\documentclass[12pt]{article}
\usepackage{pic03}
\usepackage{hyperref}
\usepackage{graphicx}

\begin{document}

\title{\bf HEAVY QUARKONIA SPECTROSCOPY AND DECAYS} 
\author{
Hanna Mahlke-Kr\"uger \\
{\em Cornell University, Ithaca, NY 14853}}
\maketitle

%
%
\begin{figure}[h]
\begin{center}
%
%
%
%
\vspace{4.5cm}
\end{center}
\end{figure}

\baselineskip=14.5pt
\begin{abstract}
The experimental status of stable bound states made out of heavy
quarks is reviewed. 
The need for a way to deal with the non-perturbative transitions
involved calls for precision measurements on one hand, and for
discovery of as yet undetected states to confirm predictions
on the other.
In this article, recent experimental contributions to heavy
quarkonia spectroscopy and decay will be reviewed, mostly 
from data analyzed by the BES~and CLEO~collaborations.

The most dramatic recent discoveries include the detection of the first
stable $L=2$ state, and a first non-pionic hadronic transition
in the $\Upsilon$ system, and a first measurement of
$\chi_{cJ} \to \Lambda \bar \Lambda$. 
Scans of the $\psi'$ and $\Upsilon(1,2,3S)$ resonances
are used to add information on partial and total decay widths.
 
\end{abstract}
\newpage

\baselineskip=17pt

\section{Heavy Quarkonia}
\begin{figure}[htbp]
\vspace{-1.0cm}
\hspace{2.5cm}
\includegraphics[width=9cm]{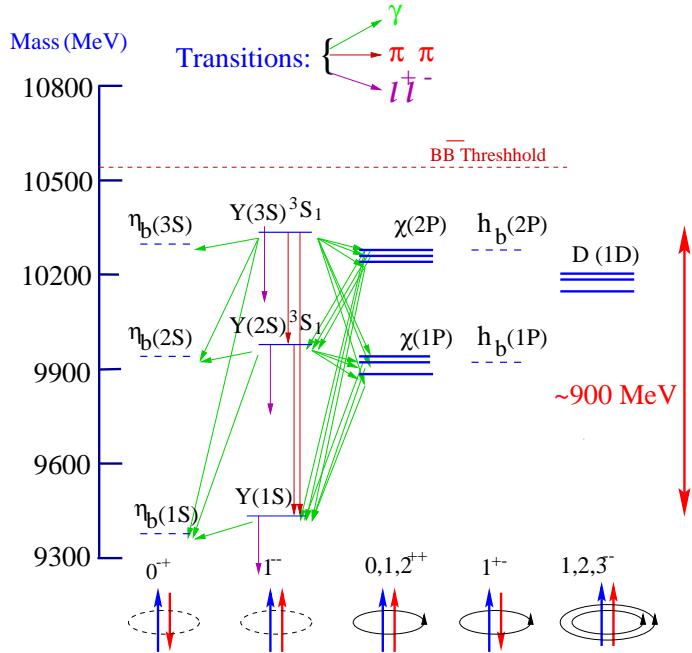}
 \caption{\it The spectrum of stable $b\bar b$ states and allowed
transitions within the system.\newline
Parallel arrows at the bottom stand for a symmetric configuation,
producing a state of total spin $S=1$,
antiparallel arrows for an antisymmetric one ($S=0$). Also
indicated is the orbital momentum, increasing from left to right.
A similar spectrum exists for charmonium. The states are labelled
similarly, albeit with a subscript $c$ instead of $b$, and the
\ $^{2S+1}L_J=^3S_1$ states are called $\psi$ ($J/\psi$ for $n=1$). 
    \label{fig:bottomonium} 
}
\end{figure}

Heavy quarkonia are non-relativistic 
strongly bound quark-antiquark states.
Owing to asymptotic freedom, non-relativistic quantum mechanics
should hold apply to a good approximation to heavy
$Q \bar Q$ systems~\cite{AppelquistandPolitzer}.
This allows us to
compare heavy quarkonia to another bound particle-antiparticle
state with slowly moving constituents, namely positronium.
In fact, a similar spectrum of
bound states is expected; see Figure~\ref{fig:bottomonium}.
The states are characterized by the following quantum
numbers: radial excitation~$n$, total spin~$S$, relative 
orbital momentum~$L$, total spin 
$\vec J = \vec L + \vec S$, parity~$P$, and $C$.
Not all states have been verified by experiment.
Just as positronium, held together electromagnetically, 
provided a testing ground to gain understanding about 
that underlying force, studing heavy quarkonia allows us to
take a better look at QCD. 

The large heavy quark masses result in small values of the running
strong coupling constant in annihilation and production processes.
In contrast to this, transitions between $Q\bar Q$ states, such as 
radiative de-excitation or
splitting off a gluon pair that turns into a pion pair, thereby
producing a lower-lying state, 
are soft processes. 
Thus many of the processes belong to the regime of 
non-perturbative~QCD, a region that lacks thorough theoretical 
understanding as of now.

The strong interaction has an impact on
many measurements investigating weak physics as well.
Also, new physics can turn out to be strongly coupled.
This means that any opportunity that presents itself to study
aspects of~QCD as an example of a strongly coupled theory must be 
used.

The bound $t \bar t$ state is very short lived since 
it can decay via the weak force, which is not possible
for charmonium and bottomonium. This makes it unsuitable
for studies of the strong interaction.
Also, if the $c\bar c$/$b \bar b$ states have masses above open 
their respective flavor production
threshold ($B\bar B$ for $b \bar b$ states and $D \bar D$
states for $c \bar c$ systems), they decay fairly quickly.
This leaves eight quasi-stable states in charmonium and thirty
in bottomonium for spectroscopy studies.

\subsection{Theoretical Understanding}
As mentioned above, heavy onia spectroscopy highlights the soft
regime of~QCD, making it impossible to use perturbative methods
to calculate transition quantities. Perturbation theory does 
describe the long distance part of the heavy quark potential.

This puts emphasis on the need for calculation methods 
that can handle non-perturbative phenomena such as Lattice QCD.
Recent developments allow to overcome the previous limitations 
that prevented unquenched calculations, {\it i.e.}~that did not treat
light quark loops, a neglect that was estimated to easily 
contribute uncertainties at the $10\%$~level. As these become
available, it is essential that their results be subject to
critical comparison with experiment in order to enhance
confidence before employing the methods to other areas such
as the weak sector of heavy quark physics.

Another solution is to use the quark velocity and $\alpha_s$ 
as expansion parameters, opening the way for effective theories
of the strong interaction. 

Yet another approach is the use of phenomenological models, 
the parameters
of which must be determined from experimental data. 
A common ansatz is a potential 
consisting of a Coulomb like term $\sim 1/r$, where $r$ is the
distance between the quarks, and a confinement term,
linear in~$r$.

\vfill

\section{Production}
\begin{figure}
\hspace{2cm}
\includegraphics[angle=270,width=6cm]{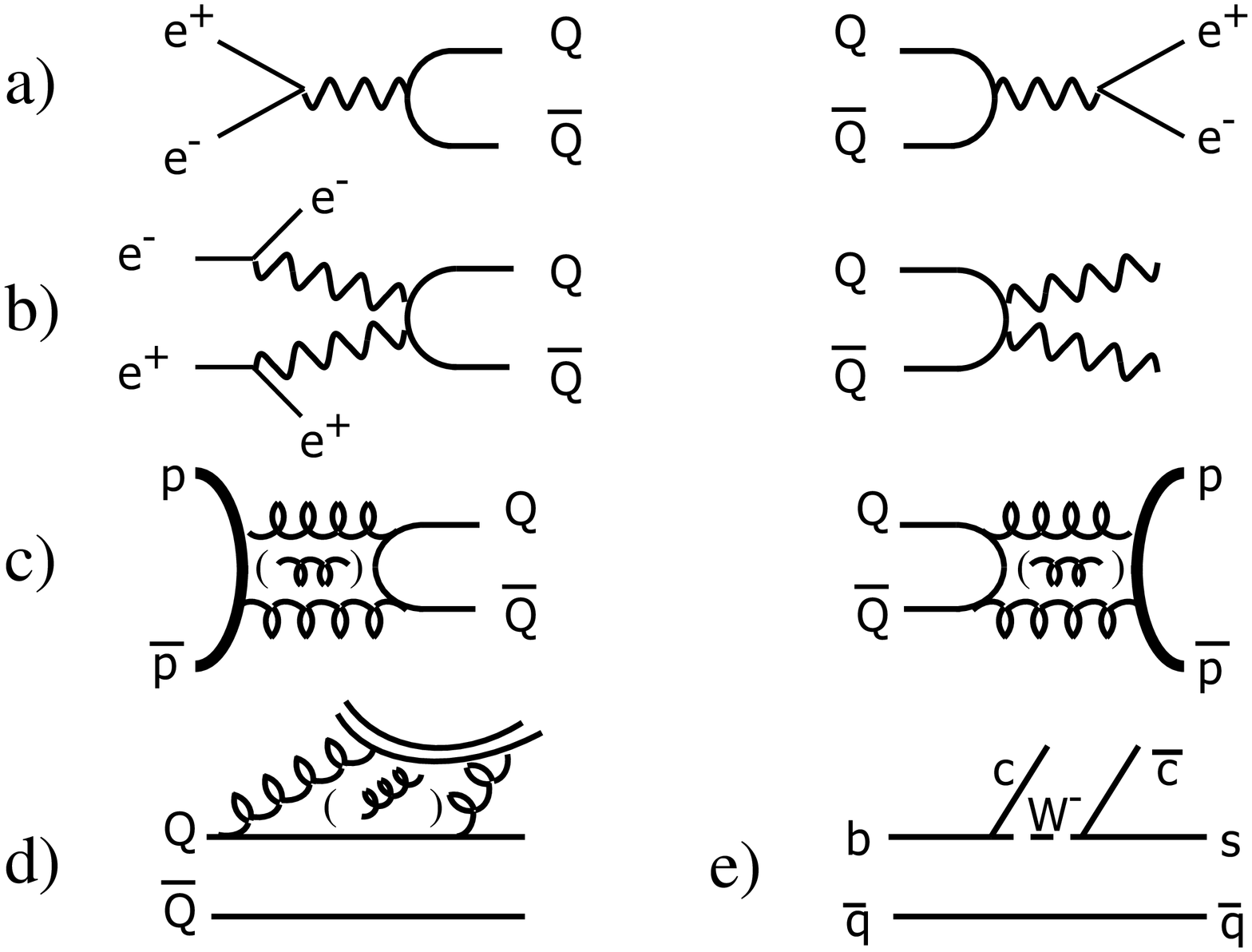}
 \caption{\it 
    Heavy quarkonia production diagrams.\newline
    Production (left) and their corresponding decay (right) 
    processes: \newline
    a)~$e^+e^- \to \gamma^* \to Q \bar Q$,
    b)~$\gamma\gamma \to Q \bar Q$,
    c)~$p \bar p \to gluons \to Q \bar Q$.\newline
    d)~Quarkonium de-excitation by emission of two pions, 
    e)~creating charmonium from a $B$~meson.
    \label{fig:FeynDiags} 
}
\end{figure}

Quarkonia can be produced in several ways, which reach
different states within the spectrum.
The first three listed here are mere reversals of 
$Q \bar Q$~decay processes and are sketched in
Figure~\ref{fig:FeynDiags}a), b), and c).

In electron positron
colliders, the reaction $e^+ e^- \to \gamma^* \to q\bar q$
results in states that can couple to a virtual photon,
namely $n^3S_1$ such as $J/\psi$ and $\Upsilon$ with a tiny
admixture of $n^3D_1$. Direct resonance production offers
the advantage of large production rates, giving access to
branching fractions even as small as $10^{-5}$.

Two-photon collisions allow direct
creation of $J=0,2$ states, {\it e.g.} $\eta_{[c,b]}$,
$\chi_{[c,b][0,2]}$. While they are readily available at
$e^+e^-$~machines, they suffer from small production rates. 
Still they provide an important contribution in that they
can be used for discovery purposes. 

Hadron machines, being able to produce any onia state
in principle by annihilation of the $p\bar p$~pair into
gluons, continue to contribute mostly to the study of
production of charmonia. 
This environment suffers from large background; thereby
one has to focus on exclusive decays.

Two more scenarios:
Downward transitions within the system provide an important 
route to otherwise not reachable states.
Any collider that can produce $B$~mesons,
be it a hadron accelerator or an $e^+e^-$~machine running
on the $\Upsilon(4S)$~resonance, has access to $c\bar c$~states
through weak decays of the $b$~quark. 
These two processes are sketched in~Figure~\ref{fig:FeynDiags}~d) and~e).

An important background for the reaction $e^+e^- \to Q\bar Q \to X$, 
or more explicitly, 
$e^+e^- \to \gamma^* \to Q\bar Q \to \gamma^* \to X$, is the case
in which  
no intermediate $Q \bar Q$~resonance is formed. The presence of this
channel adds to the measured cross-section both directly and by 
interference, which can be a sizeable 
contribution~\cite{cont_interference}.
In most measurements, this contribution is not taken into account 
or subtracted.
This background needs to be either measured, by running off the 
relevant resonance, or calculated. 
In measurements of the cross-section as function of
energy (scans), the non-resonant production can be explicitly taken into
account when fitting the line shape. 

\section{Transitions}

\subsection{Hadronic Transitions}
In order to conserve charge, transitions can either happen
by emitting neutral particles or a charged pion pair.
Single $\pi^0$ transitions are isospin suppressed. 
Kaon pair transitions are phase space prohibited.

\subsubsection{Non-Dipion transitions}
\begin{figure}[htbp]
  \centerline{\hbox{ 
    \includegraphics[bb=0 200 535 688,width=6.0cm]{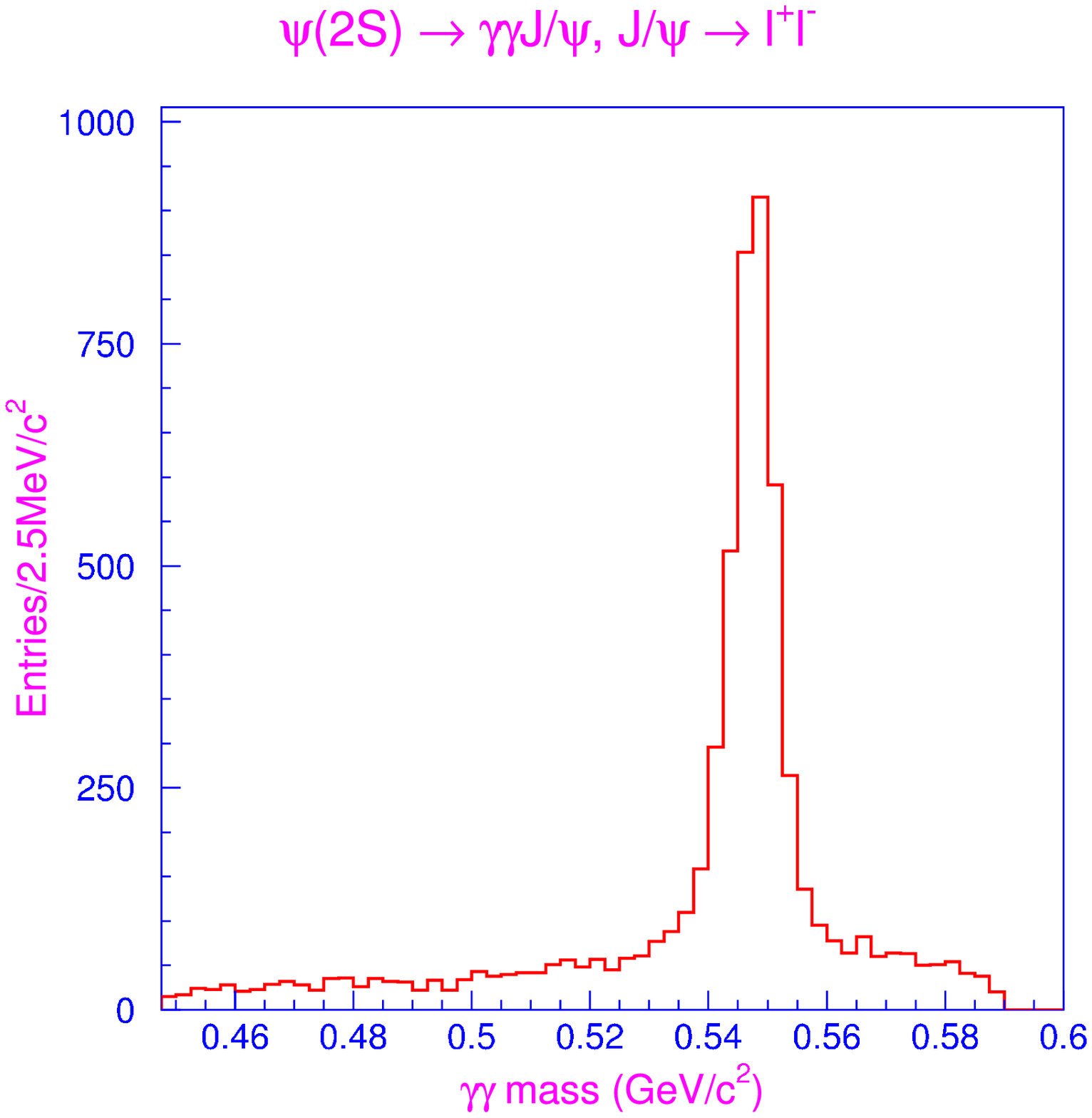}
    \includegraphics[bb=0 200 535 688,width=6.0cm]{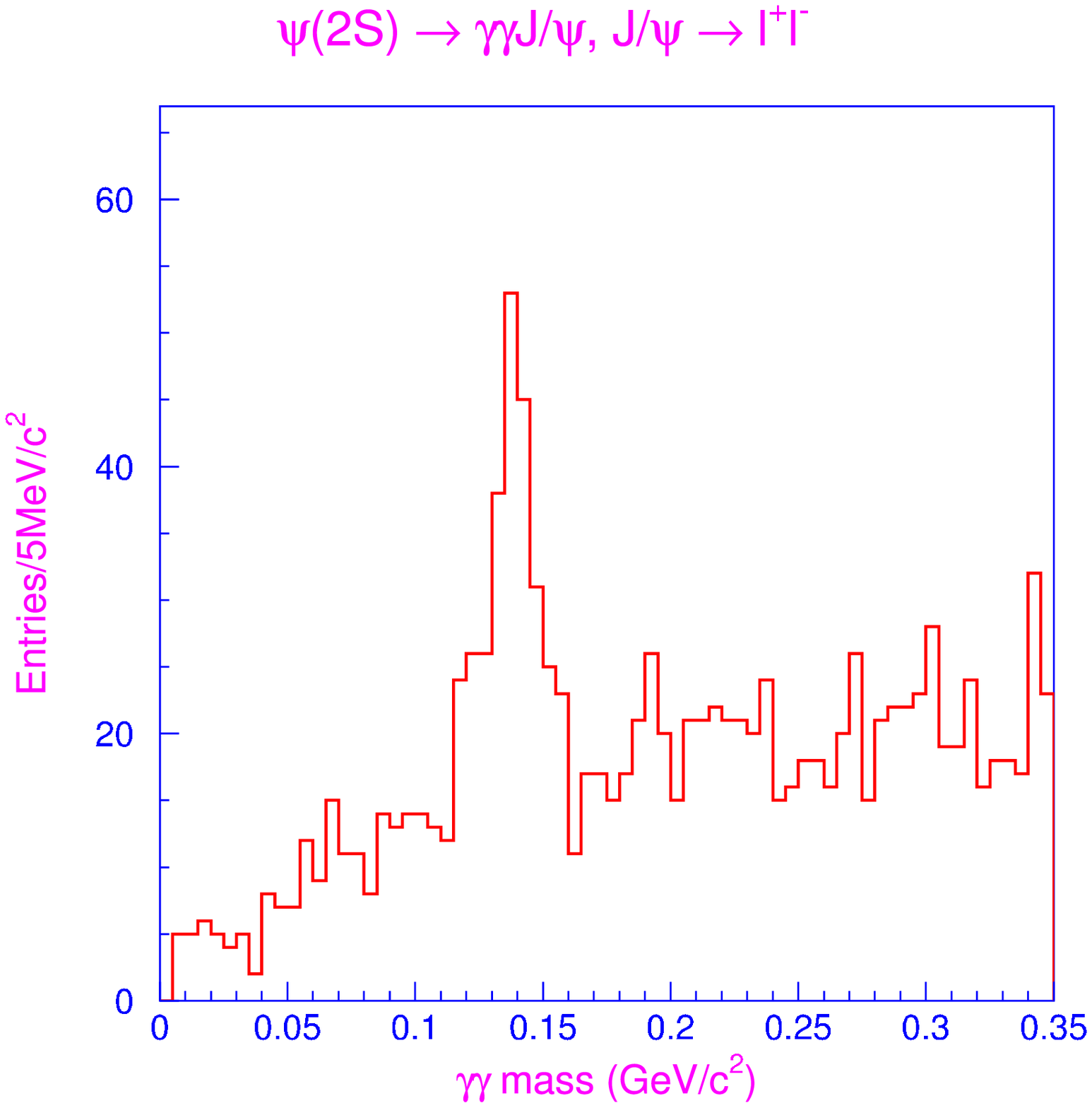}
    }
  }
 \caption{\it
       Evidence for $\pi^0$ and $\eta$ transitions 
       in charmonium~\cite{CernCourier}.
    \label{psiprimerare} }
\end{figure}

By studying the decay $\psi' \to J/\psi \gamma \gamma$
and plotting the invariant mass of the two photons,
evidence for a single pion and $\eta$ transitions can be 
obtained. 
The BES Collaboration, using their 15M~$\psi'$ sample, will
be able to perform precise measurements of these transitions.
To illustrate the quality of the data sample, Figure~\ref{psiprimerare}
displays the result of a study of the decay 
$\psi' \to \gamma \gamma l^+ l^-$,
where clear $\pi^0$ and $\eta$
signals are observed in the distribution of the
invariant $\gamma\gamma$~mass~\cite{CernCourier}. 
 
\begin{figure}[htbp]
  \centerline{\hbox{ \hspace{0.2cm}
    \includegraphics[width=6.5cm]{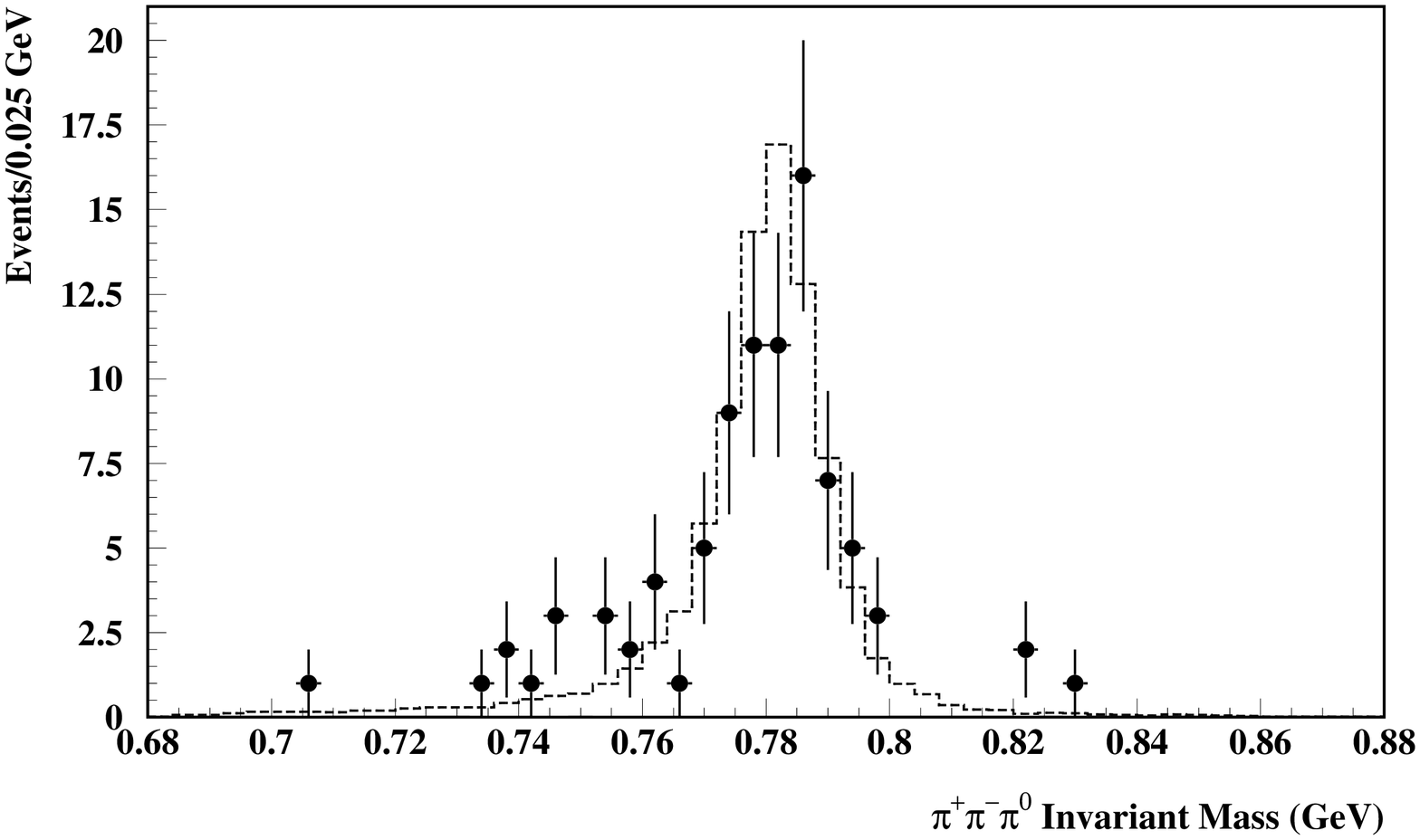}
    \hspace{0.3cm}
    \includegraphics[width=6.5cm]{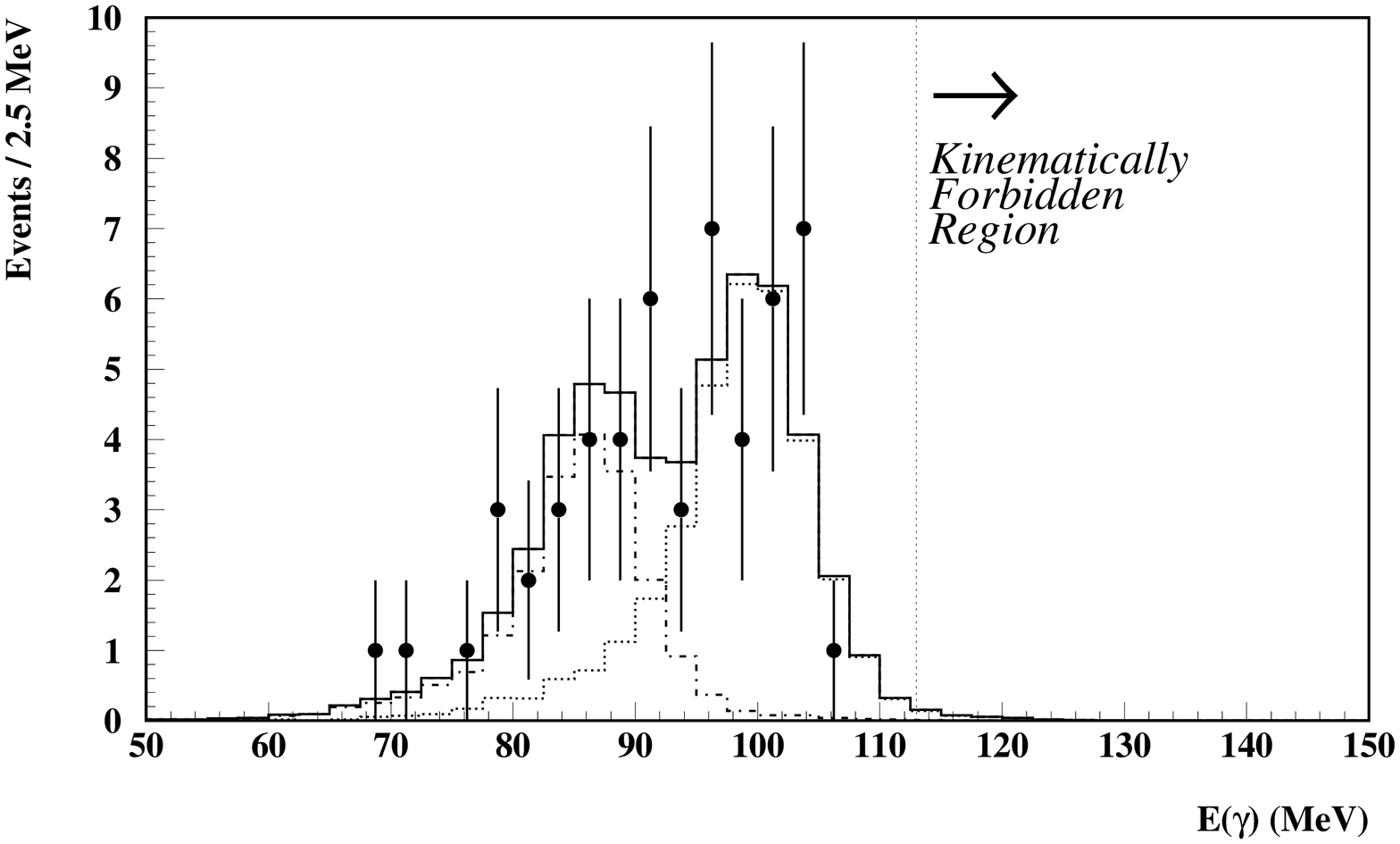}
    }
  }
 \caption{\it
    First observation of a non-pionic hadronic transition in bottomonium:
    Left: The three pion invariant mass, indicating the presence of
    an~$\omega$.
    Right: Fit to the photon energy spectrum, with the individual contributions 
    from $\Upsilon(3S) \to \gamma \chi_{b1}'$ (dotted) and
    $\Upsilon(3S) \to \gamma \chi_{b2}'$ (dash-dotted) overlaid.
    \label{fig:omegatransition} }
\end{figure}

The first non-pionic hadronic transition
in the $\Upsilon$ system was recently found by the 
CLEO~Collaboration. In their sample of 
$(5.81\pm 0.21)\times 10^{6}$ $\Upsilon(3S)$~decays
evidence was found for the decay chain
$\Upsilon(3S) \to \gamma \chi_{b1,2}' \to
\gamma \omega \Upsilon(1S)$ with $\omega \to \pi^+ \pi^- \pi^0$
and $\Upsilon(1S)\to l^+ l^-$. 
(The decay $\chi_{b0}' \to \omega \Upsilon(1S)$
is phase space suppressed.) Requiring the presence of an
$\Upsilon(1S)$~candidate, identified through its decay into
a high-momentum lepton pair, guarantees that 
the background from $udsc$~pair production in the data sample
is negligible. 
The measured branching fractions, obtained by maximum likelihood
fit to the energy spectrum of photons not assigned to the
$\pi^0$~candidates (Figure~\ref{fig:omegatransition}), are at the 
percent level~\cite{omegatrans},
thereby confirming the prediction that such transitions
should have sizeable rates. 
 
\subsubsection{Dipion transitions}
\begin{figure}[htbp]
  \centerline{\hbox{ \hspace{0.2cm}
    \includegraphics[width=6.5cm]{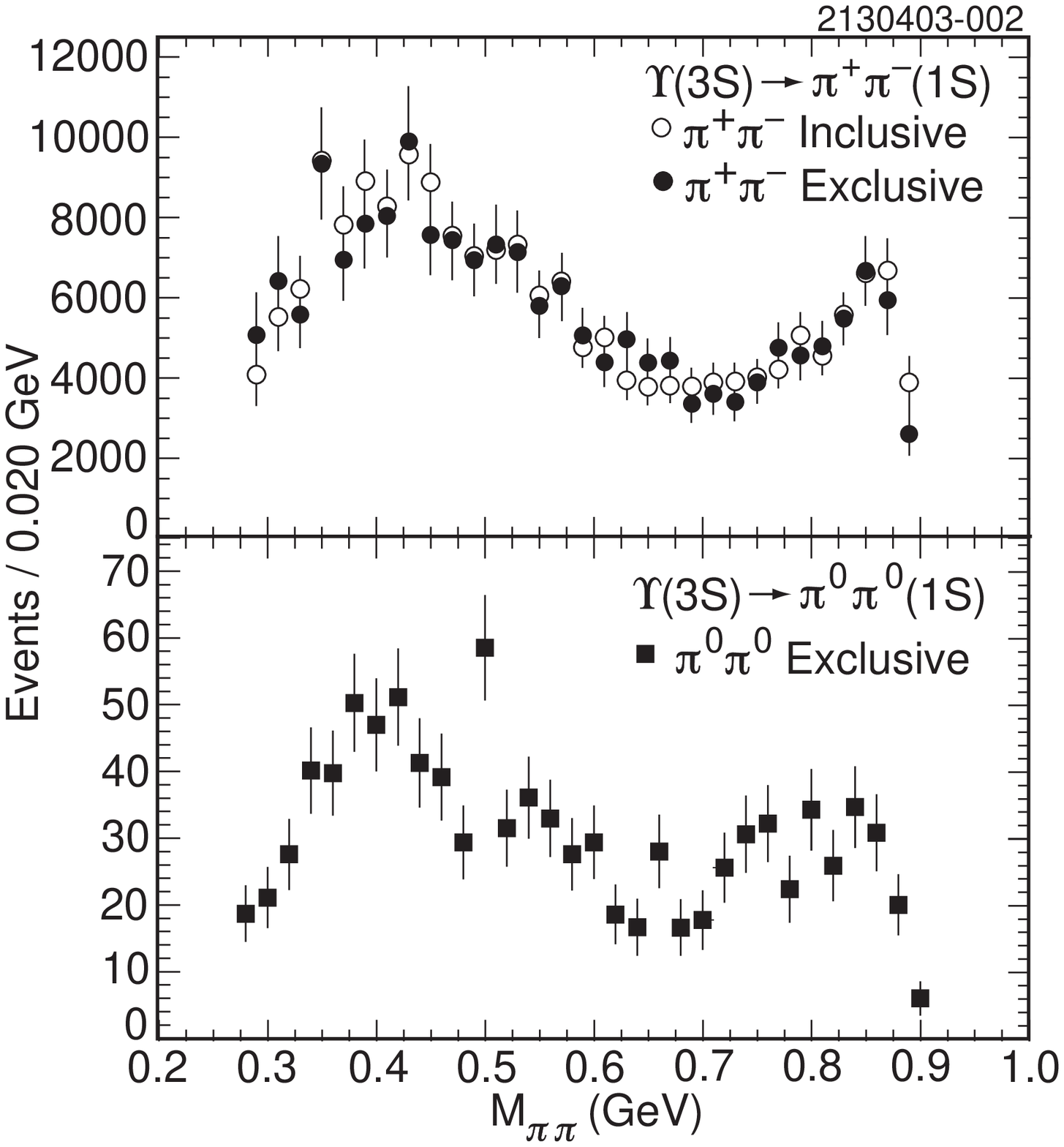}
    \hspace{0.3cm}
    \includegraphics[width=6.5cm]{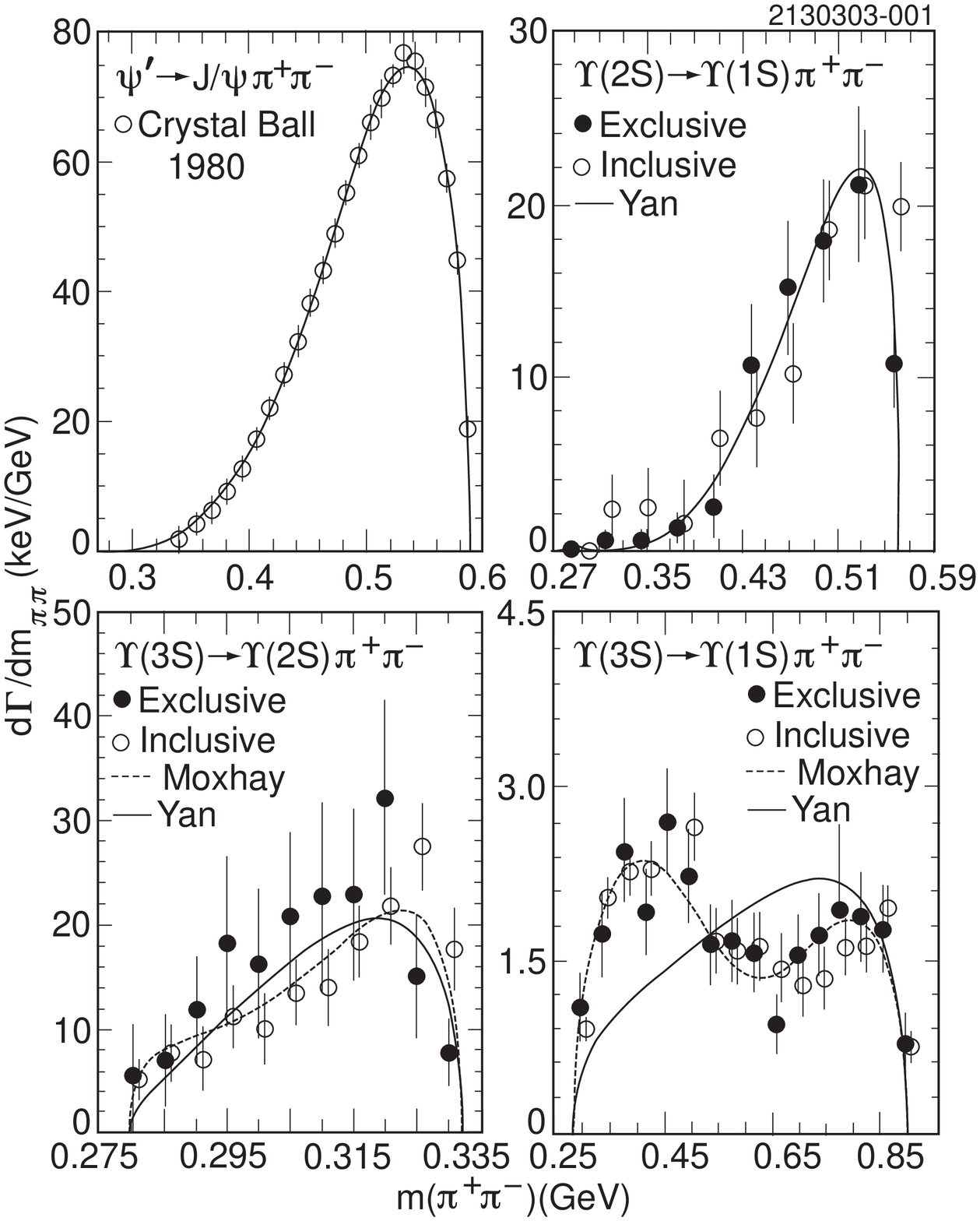}
    }
  }
 \caption{\it
     The dipion invariant mass spectrum in hadronic onia transitions.
     Left: new preliminary CLEO data,
     right: previous results from
     Crystal Ball~\cite{cbalpsidipion} 
     ($\psi' \to \psi \pi^+ \pi^-$)
     and CLEO~\cite{cleo_dipion} 
      ($\Upsilon(2S) \to \Upsilon(1S) \pi^+ \pi^-$, 
       $\Upsilon(3S) \to \Upsilon(2S) \pi^+ \pi^-$, and
       $\Upsilon(3S) \to \Upsilon(1S) \pi^+ \pi^-$).
    \label{fig:dipion} }
\end{figure}

When comparing the dipion invariant mass spectrum in
$\psi' \to \psi \pi^+ \pi^-$, 
$\Upsilon(2S) \to \Upsilon(1S) \pi^+ \pi^-$, 
$\Upsilon(3S) \to \Upsilon(2S) \pi^+ \pi^-$, and
$\Upsilon(3S) \to \Upsilon(1S) \pi^+ \pi^-$, 
the last has a distinctive double-peak structure
(Figure~\ref{fig:dipion}, right)~\cite{cleo_dipion}. 
Many models have been developed that try and fit this
behaviour, albeit lack of precision did not allow
a clear discrimination between the models.
CLEO has presented new preliminary data~\cite{cleo_aps}. 
The structure is
confirmed in both $\pi^+\pi^-$ and $\pi^0\pi^0$
reactions, measured exclusively and inclusively, 
see Figure~\ref{fig:dipion}.

\section{Charmonium Singlet States}
The ground state of charmonium, $\eta_c$, has been
studied for some time. Theoretical predictions
of the mass and width of its first radial excitation have so 
far been limited to potential model calculations and
based on drawing an analogy of the $S=0$ pair ($\eta_c',\ \eta_c$)
to the $S=1$ pair ($\psi',\ J/\psi$).
An early measurement by Crystal Ball yielded a particularly
low $\eta_c'$~mass, thereby introducing a lower value for
the mass splitting $\Delta m = m(\eta_c') - m(\eta_c)$.
Recent measurements of CLEO and BaBar using two photon 
events and Belle studying $B\to K \eta_c' \to K K_s K^- \pi^+$ 
decays~\cite{etacprime_twophoton} 
give ample indication towards a higher value, 
as does a Belle measurement of the spectrum of the
system recoiling against the $J/\psi$~\cite{etacprime_recoil} in
$e^+ e^- \to J/\psi X$. At the same time, the $\eta_c$~mass
has been remeasured in these experiments. A precision
measurement was performed by BES using radiative decays in
a 58M~$J/\psi$~data set~\cite{besetac2003}.
A summary of experimental data on the $\eta_c$~and $\eta_c'$~masses 
can be found in Table~\ref{tab:etacprime}. 
Of particular interest to theorists will be the mass splitting
between the two $n=1,\ L=0$ states, $m(\eta_c')-m(\psi')$,
in comparison to the $n=0$ states, used to calculate the strength
of the spin-spin interaction term in non-relativistic potential
models. 
 
Many attempts have been made to measure the $1^1P_1$
state in charmonium, labeled $h_c$~\cite{h_csearches}. 
So far, the results are suggestive, but not conclusive. 
In \cite{KuangTuanYan}, the branching fraction 
${\cal B}(\psi' \to \pi^0 h_c)$
is predicted to be as large as $3.7\times 10^{-3}$,
with ${\cal B}(h_c \to \gamma \eta_c)$ as large as
$50\%$. Combined with a reasonable detection efficiency, 
even a moderate size $\psi'$ data sample should 
be able to test this prediction. 
  
\begin{table}
\centering
\caption{ \it Experimental $\eta_c$ and $\eta_c'$ mass measurements.
}
\vskip 0.1 in
\begin{tabular}{|l|c|c|} \hline
experiment & $m(\eta_c)\ (MeV)$ & $m(\eta_c')\ (MeV)$ \\
\hline
\hline
 Crystal Ball & $2980 \pm 8$~\cite{etac_cbal}
                                 & $3594 \pm 5 $~\cite{etacprime_cbal}\\ 
 Belle 2003 (exclusive) 
            & $2979 \pm 2(stat)$ & $3654 \pm 6(stat) \pm 8(syst) $ \\ 
 Belle 2003 (recoil) & $2962 \pm 13 (stat) $ & $3622 \pm 12 (stat)$ \\ 
 BaBar 2003 ($\gamma\gamma$)
            & $2983.3 \pm 1.2(stat) \pm 1.8(syst)$ 
            & $3632.2 \pm 5.0 (stat) \pm 1.8 (syst)$ \\ 
 CLEOII/II.V 2003 ($\gamma\gamma$)
            & $2984.7 \pm 2.1 $ & $3642.7 \pm 4.1 \pm 4.0(syst)$ \\ 
 BES 2003 ($J/\psi$ decay)  
            & $2977.5 \pm 1.0(stat) \pm 1.2(syst)$ & \\
\hline
 Potential models~\cite{etacprime_models} & & $3594 - 3629$ \\
\hline
\end{tabular}
\label{tab:etacprime}
\end{table}

\section{Search For New States in Bottomonium 
Using Radiative $\Upsilon(nS)$ decays}
In contrast to the situation in charmonium, no singlet 
$b \bar b$ state has been observed yet:
the bottomonium ground state $1^1S_0$, $\eta_b(1S)$,
has not been seen (and neither have its excitations),
which is also true for the $h_b$ states, $n^1P_1$. Since
they are not accessible in direct production, they
are searched for by making use of photon 
transitions. 

CLEO studied magnetic dipole (M1)~transitions from the 
triplet~$S$ states\footnote{
These allow $\Delta L = 0$. Direct transitions have $\Delta n = 0$,
whereas the ones with $\Delta n > 0$ are called hindered.
Hindered transitions, due to small wave function overlap
made possible only by relativistic corrections, are
suppressed, but benefit from a $E^3$ dependence in the
transition rate, where $E$ is the energy difference
between the states. 
}, using $4\times 10^6$ $\Upsilon(3S)$ and
$3\times 10^6$ $\Upsilon(2S)$ decays.  
The reactions examined in an inclusive
photon spectrum analysis are:
$\Upsilon (3S) \to \eta_b(2S)\gamma$, 
$\Upsilon (3S) \to \eta_b(1S)\gamma$, 
$\Upsilon (2S) \to \eta_b(1S)\gamma$, 
and 
$\Upsilon (3S) \to h_b(1P)\pi\pi$ with $h_b \to \gamma \eta_b$. 
A search window was defined based on predictions of the
hyperfine splitting between the $\eta_b(nS)$ and $\Upsilon(nS)$
states and the $h_b$ and $\chi_b$ states. 
As a preliminary result~\cite{cleo_aps}, no
signal was found, but upper limits at the $10^{-3}$
level in the corresponding search windows have been
set on the branching fractions. This already rules out some
of the models.
Also, it states that a significant improvement in data
sample size will be needed to establish a signal.

\subsection{$L=2$ in Bottomonium }
The $\Upsilon(1D)$, a $L=2$ state, is unique in that it
is the only stable high-$L$ state in heavy quarkonia as
all others lie above open production threshold. Providing
experimental evicendence is not only a matter of principle,
but allows for discrimination amongst various theoretical
models that were tuned on $L=0,1$ states. A second way in
which such a state is special is that it
decays preferentially electromagnetically, thereby
is comparatively narrow. It was searched for by the
CLEO Collaboration in 
$\Upsilon(3S)$~decays in four-photon cascades: 
$\Upsilon(3S) \to \gamma \chi_b(2P)$, 
$\chi_b(2S) \to \gamma \Upsilon(1D)$, 
$\Upsilon(1D) \to \gamma \chi_b(1P)$, 
$\chi_b(1P) \to \gamma \Upsilon(1S)$. 
A signal of $6.8\sigma$~significance is seen at
$10161.2\pm 1.6$~MeV\cite{ups1dleppho}, 
inconsistent with $J=3$.
Theory predicts dominance of the $J=2$ state over
$J=1$ by a factor of about six~\cite{GodfreyRosner}. 
The new state is therefore assigned to be the 
$\Upsilon(1^3D_2)$.

\section{Scans}
Precise cross-section determinations around a resonance 
as function of the center-of-mass energy, often referred 
to as ''scans'', are an excellent tool to determine
the resonance parameters with as little bias as
possible.  

\subsection{BES $\psi'$ scan}
\label{sec:BESscan}
The BES Collaboration has determined $\psi'$ resonance
parameters~\cite{psiprimescan}, 
in particular studying the reactions
$\psi' \to $~hadrons, $\psi' \to \pi^+ \pi^- J/\psi$,
and $\psi' \to \mu\mu$. The quantities determined in
a simultaneous fit to these cross-sections
were $\Gamma_{tot}$, $\Gamma_{\mu}$,
and $\Gamma_{J/\psi \pi^+ \pi^-}$, the derived quantities
are $\Gamma_{had}$, ${\cal B}_{had}$, ${\cal B}_{l}$, and
${\cal B}_{J/\psi \pi^+ \pi^-}$:
\begin{itemize}
\item
The channel $\psi' \to \mu\mu$ can be combined with
other leptonic width measurements, testing the sequential
lepton hypothesis, which states that : 
$$
\frac{B_l}{v_l (\frac 3 2  - \frac 1 2 v_l^2) } \mbox{ with }
v_l= \sqrt{1- 4m_l^2/M_{\psi'}^2}
$$
should be the same number for $l=e,\mu,
\tau$. The denominator is about unity for
$l=e,\mu$ and $0.39$ for $l=\tau$. 
This BES determination of the muonic branching
fraction yielded 
${\cal B}_{\mu}=(9.2 \pm 0.8)\times 10^{-3}$.
The E760~Collaboration published a value of ${\cal B}_e =
(8.3 \pm 0.5 (stat) \pm 0.7 (syst))\times 10^{-3}$. 
A separate (and first direct) measurement of ${\cal B}_{\tau}$
was performed as well by the BES~collaboration~\cite{Btau_BES}, 
resulting in a
value of ${\cal B}_{\tau}=(2.71 \pm 0.43 \pm 0.55)\times 10^{-3}$.
After applying the above correction factor to the
the $\tau$ partial decay width, it can be compared with the
other leptonic branching fractions. Agreement within experimental
errors is observed.
\item
The $\psi' \to \pi^+ \pi^- J/\psi$
line shape is of importance since this decay is frequently 
used as a normalizing mode. The precision obtained on the
branching fraction in this
measurement is 4.4\%, improved over the PDG2002 value of~5.2\%.
\item
Finally, using the relations
$\Gamma_{tot} = \Gamma_{had} + \Gamma_{\mu} + \Gamma_{e}
+ \Gamma_{\tau}$ and 
$\Gamma_{e} = \Gamma_{\mu} = \Gamma_{\tau}/0.39$
and measuring the $\psi' \to $~hadrons 
cross section, $\Gamma_{tot}$ was obtained as
\linebreak
$\Gamma_{tot}^{\psi'} = (264 \pm 27)$keV. 
This can be compared with a PDG2002 value of 
$\Gamma_{tot}^{\psi'}=(300 \pm 25)$keV~\cite{pdg2002} 
and a $p \bar p$ scan result by E760 of 
$\Gamma_{tot}^{\psi'}=(306 \pm 36 \pm 16)\,$keV~\cite{gammatot_psi2s_e760}.
\end{itemize}

\subsection{CLEO $\Upsilon(1,2,3S)$ scans}
CLEO hopes to improve the precision of the
leptonic width $\Gamma_{e}$
from currently $2,~4,~9\%$ for $\Upsilon(1,2,3S)$
down to the level of $2\%$ for each of the three.
One reason is that this parameter enters many
other measurements. In addition, it provides
a high-precision test for Lattice~QCD, which
begins to be able to reach this level of accuracy.
Preliminary results show a statistical precision
of $0.1,~0.3,~0.5\%$. Systematic errors are
still being evaluated.

\section{Decays}
Heavy onia can decay via the electromagnetic or
the strong force. 
Possible decay mechanisms
for a heavy onium state are annihilation of the two
heavy quarks into two leptons, one or more photons, or
two or three gluons.
For states below open flavour production
threshold, the decay into leptons only contributes little
to the total width ($12\%$~for the $J/\psi$, less
for other states), and the remaining hadronic
rate is by far not accounted for by the exlusive
decays measured so far. 
To some, the only pieces of information
of interest are the leptonic decay widths, since the rest has
to consist of radiative or hadronic decays. How exactly
this greater remainder of the total rate is divided
up into specific final states is largely governed by
fragmentation dynamics.
Electromagnetic decays have been discussed
in Section~\ref{sec:BESscan}.

\subsection{First Evidence of $\chi_{cJ} \to \Lambda \bar \Lambda$ }
BES has reported the observation of 
$\chi_{cJ} \to \Lambda \bar \Lambda$~\cite{chicjtolambdalambdabar}. 
Besides this being the first measurement of the branching fraction,
this decay is of interest in comparison to 
$\chi_{cJ} \to p \bar p$. 
It has been shown that the lowest
Fock-state expansion of charmonium states (``color singlet model'', 
``CSM'') is insufficient to describe $P$-wave quarkonium decays,
both inclusively and exclusively,
and that use of the next higher Fock-state
 (``color octet mechanism'', ``COM'') 
improves the agreement with experiment.
The agreement of COM based prediction with 
the total measured width of the $\chi_{c0}$ as well as that for the
partial width of $\chi_{cJ} \to p \bar p$, obtained by using
a carefully tuned nucleon wave function~\cite{chipp_wf}, was encouraging. 
Generalizing the nucleon wave function to other baryons lead to 
a prediction for the partial width of $\chi_{cJ} \to \Lambda \bar \Lambda$
as being about half of that for $\chi_{cJ} \to p \bar p$~\cite{chipp_wf}
for $J=1,2$. 

\begin{figure}[htb]
\begin{center}
\includegraphics[width=8cm,height=5cm]{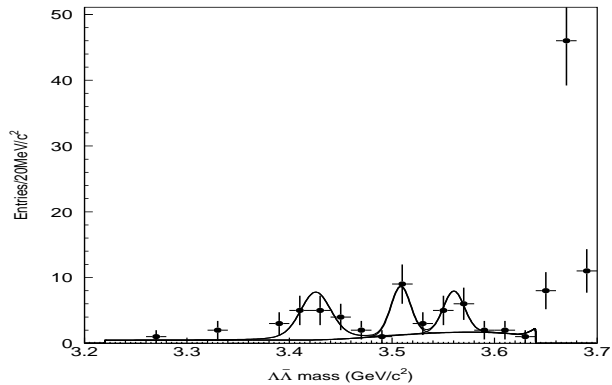}
\end{center}
 \caption{\it
      \vspace*{-5mm}
      Distribution of the invariant 
      $\Lambda \bar \Lambda$ mass
      in $\psi' \to \gamma (p \pi^-) (\bar p \pi^+)$
      events~\cite{chicjtolambdalambdabar}.
    \label{fig:chicpeaks} }
\end{figure}

The analysis uses a recently collected $15M$ sample of $\psi'$~events, 
decaying radiatively to the $\chi_c$ states. 
The desired channel is identified
through one $\pi^- p$ and one charge conjugated candidate. The distribution
of the resulting $\Lambda \bar \Lambda$ events is shown in 
Figure~\ref{fig:chicpeaks}, overlaid with the fit result, with the masses
of the three $\chi_c$ states as fit parameters. 
As a normalizing channel, $\psi' \to \pi\pi J/\psi$ is used. 
The branching fractions thus derived
are listed in Table~\ref{tab:chictolambdalambdabar}, together with the
result for $\chi_{cJ} \to p \bar p$ from~\cite{pdg2002}.
The large experimental errors do not justify ruling out the prediction. 

\begin{table}
\centering
\caption{ \it Experimental results on $\chi_{cJ} \to \Lambda \bar \Lambda$,
and comparison with $\chi_{cJ} \to p \bar p$.
}
\vskip 0.1 in
\begin{tabular}{|l|c|c|} \hline
\rule[-1.5mm]{0mm}{5.5mm}
& ${\cal B}(\chi_{cJ} \to \Lambda \bar \Lambda)$ in $10^{-4}$ & 
${\cal B}(\chi_{cJ} \to p \bar p)$ in $10^{-4}$ \\
\hline
$J=0$ & $ 4.7^{+1.3}_{-1.2} \pm 1.0 $ 
      & $ 2.2 \pm 0.5 $ \\
$J=1$ & $ 2.6^{+1.0}_{-0.9} \pm 0.6 $ 
      & $ 0.7 \pm 0.3 $ \\
$J=2$ & $ 3.3^{+1.5}_{-1.3} \pm 0.7 $ 
      & $ 0.7 \pm 0.1 $ \\
\hline
\end{tabular}
\label{tab:chictolambdalambdabar}
\end{table}

\subsection{Hadronic Decays}
Predictions exist that relate the branching fraction
for the $1^3S_1$ into hadronic final states to those
of its first radial excitation: if this decay happens
predominantly via annihilation into gluons, then
the wave function overlap should be the only only
difference between the decay of the two states
(aside from the difference in center-of-mass
energy that the gluons have, which is not vastly different). 
This in turn can be
taken from the leptonic branching fraction, another
annihilation process. Ignoring the running of the strong 
coupling constant\footnote{
A value of 0.85 is quoted for 
$\left(\alpha_s(m_{\psi(2S)}) / \alpha_s(m_{J/\psi}) \right)^3$
in~\cite{guandli}.
} $\alpha_s$, one obtains
\begin{equation}  
Q=\frac{{\cal B}(\psi(2S) \to H )}{{\cal B}(J/\psi \to H)}   
\approx
\frac{{\cal B}(\psi(2S) \to e^+ e^- )}{{\cal B}(J/\psi \to e^+ e^-)}.   
\label{eqn:Q}  
\end{equation}
Using the leptonic branching fractions  
${\cal B}(J/\psi \to e^+ e^-) = (5.93 \pm 0.1 ) \times 10^{-2}$  
and
${\cal B}(\psi(2S) \to e^+ e^-) = 
(7.3 \pm 0.4) \times 10^{-3}$~\cite{pdg2002},  
the expected value for the ratio is $Q=(12.3 \pm 0.9)\%$.
Different views exist as to whether this prediction is 
valid only for the inclusive process $\psi(nS) \to X$ or also
for specific final states.
Some theorists posit that it should hold also for radiative decays.
Moreover, a similar relation should exist for the ratio for the 
$S=0$~states. 

A number of channels have been studied in charmonium, including
decays into a vector ($V$) and a pseudoscalar ($P$) particle,
axialvector pseudoscalar,
vector plus a tensor ($T$), radiative decays, multibody decays,
as well as dibaryon final states~\cite{pdg2002, psiprimedecays}.
While many of these are not in outrageous disagreement with
the above prediction, two $VP$~modes are well-known for failing:
$\rho\pi$ ($Q_{\rho\pi}< 0.65\%$) and 
$K^* K$ ($Q_{K^* K}< 1\%$). 
Also in $VT$~states substantially lower ratios have been
observed ($Q_{\omega f_2}< 3.5\%$, $Q_{\rho a_2}< 2\%$).
The overall picture is not clear, in part due to inaccurate 
experimental results. Many a theorist has given input
on this anomaly~\cite{fourteenpercent_theory} that is often
referred to as the $\rho\pi$~puzzle, 
but so far none is able to explain all experimental results.  
It is of major interest to decide whether the $J/\psi$ rate
is enhanced or the $\psi'$ rate is suppressed.
 
In the $\Upsilon$ system, a similar relation is expected to hold. 
As in this case two excitations are below dissociation threshold,
two such ratios can be built. Using the corresponding leptonic
branching ratios, one obtains $48\%$ for $\Upsilon(2S): \Upsilon(1S)$ 
and $72\%$ for $\Upsilon(3S): \Upsilon(1S)$.  
It is by no means clear what absolute rate to expect for
$\Upsilon$~decays when extrapolating from charmonium. Depending
on the model chosen to explain the $\rho\pi$ anomaly, the rates
vary considerably. 
A preliminary CLEO result studying a variety of two-body hadronic
$\Upsilon$~decays are upper limits of $4 \cdot 10^{-5}$ or better
on $\rho\pi$, $K^*(892)\bar K$, $\rho a_2(1320)$, 
$K^*(892) \bar K_2^*(1430)$, $\omega f_2(1270)$, $b_1(1235)\pi$,
and $K_1(1400)\bar K$ on the $\Upsilon (1,2,3S)$ 
resonances~\cite{cleoupsilondecays}. In particular, it is found that 
${\cal B}(\Upsilon(1S) \to \rho \pi)$ is well below $10^{-5}$.

\section{Summary}
Heavy quarkonia continue to
provide a testing ground for QCD calculations. 

A wide variety of measurements is being
carried out, with hopes for many more results once
BES and CLEO have analyzed their recently taken
large datasets.

The goals for the future are to establish the
states that have been predicted to exist but not
observed yet and to provide precision measurement
to allow a detailed comparison with theory.

This will hopefully make it possible to gain
further insight into the non-perturbative
realm of QCD, from which many measurements in
other areas will benefit greatly.

\section{Acknowledgements}
The author is grateful to her many collaborators on CLEO 
who provided analysis results, discussion, and guidance.
Special thanks go to X. Shen, C. Yuan, and W. Li on BES.

\end{document}